\documentclass[]{spie}  

 \usepackage{dsfont}
\usepackage{graphicx}
\usepackage{subcaption}

\usepackage{amsmath,amsfonts,amssymb}
\usepackage{graphicx}
\usepackage{braket}
\usepackage[colorlinks=true, allcolors=blue]{hyperref}
\newcommand\blfootnote[1]{
    \begingroup
    \renewcommand\thefootnote{}\footnote{#1}
    \addtocounter{footnote}{-1}
    \endgroup
}

\title{Automated Polarization Basis Adjustment and Security Monitoring in Quantum Communication via Coincidence Entropies}

\author[a,b]{Tomáš Novák}
\author[a,c]{Carlos Guerra-Yánez}
\author[a]{Matěj Holubička,}
\author[a]{Josef Vojtěch}
\author[a]{Josef Blažej}
\affil[a]{Czech Technical University in Prague, Faculty of Nuclear Sciences and Physical Engineering, Department of Laser Physics and Photonics, Prague, 11519, Czech Republic}
\affil[b]{CESNET a.l.e., Generála Píky 430/26, 160 00 Praha 6, Czech Republic}
\affil[c]{Dept. of Electromagnetic Field, Czech Technical University, Technická 2, Prague, Czech Republic}

\begin{document} 
\maketitle

\begin{abstract}
Polarization-sensitive receivers for single photons are of crucial importance in various applications within the fields of quantum communication and quantum sensing, and are more commonly implemented in free-space optics rather than in optical fibers. This is primarily due to the unpredictable and varying birefringence in single-mode optical fibers. We present a method for birefringence compensation in an all-fiber detection setup that relies solely on coincidence measurements of a polarization-entangled state, or known correlations in a prepare-and-measure scenario. We define \textit{coincidence entropies} as functions of the measured coincidence counts. These quantify the randomness of measurement outcomes, remain independent of the transmitted Bell state, and serve as indicators of the degree of entanglement. By leveraging coincidence entropy as a cost function in a gradient descent algorithm, we are able to align the polarization bases between two distinct polarization-sensitive receivers. Additionally, coincidence entropies can be employed to monitor the quality of entanglement transmission, thereby enhancing the system’s ability to detect potential eavesdropping attempts, such as intercept-resend quantum attacks.



\end{abstract}


\section{INTRODUCTION}
\label{sec:intro}

In general, polarization-based transmission over single-mode optical fibers faces the primary challenge of compensating for unpredictable and time-varying unitary polarization transformations caused by fiber birefringence. These effects are mainly induced by external stress, bending, torsion, temperature fluctuations, and gradients, and become more pronounced in longer, deployed transmission lines.

Traditional approaches for polarization basis alignment in classical communication, such as polarization tracking or stabilization methods, are not applicable to entanglement distribution. This is because the measurement outcomes of a single photon from an entangled pair behave as if originating from unpolarized light. Instead, polarization correlations can only be accessed via polarization-dependent coincidence measurements.
	\blfootnote{© (2025) Society of Photo-Optical Instrumentation Engineers (SPIE). One print or electronic copy may be made for personal use only. Systematic reproduction and distribution, duplication of any material in this publication for a fee or for commercial purposes, and modification of the contents of the publication are prohibited. This is the author-prepared version.}
Several approaches for polarization basis alignment in quantum systems have been proposed. These include the use of an auxiliary laser as a reference\cite{Peranic2023}, Quantum Bit Error Rate (QBER) minimization through optimization techniques such as AI-assisted alignment\cite{Mantey2023}, stochastic optimization algorithms\cite{Shi2021}, or gradient descent\cite{Tan2024}. However, none of these methods employ a fully fiber-integrated detection setup, which may even be incompatible with some analytical techniques used in prepare-and-measure polarization-based quantum key distribution (QKD) systems\cite{Mayboroda2024}.

While constructing a detection setup using fiber-based components provides an alignment-free alternative to free-space optics, the random birefringence inherent to optical fibers must still be actively compensated for using polarization controllers. The rate at which birefringence varies limits the operational frequency of the compensation process and is primarily determined by the length of the optical fiber and its exposure to environmental variability. Importantly, the dominant source of birefringence-induced variation arises from the deployed transmission fiber, rather than the fiber-integrated detection setup, which typically resides in a static and controlled environment.

A fiber-based detection setup was implemented in\cite{Shi2021}, but it featured only a single polarization detection arm, preceded by a fast electronic polarization controller used to dynamically select the measurement basis in a prepare-and-measure scenario. In contrast, our approach employs a static selection of polarization measurement bases using a beam splitter (BS) followed by polarization beam splitters (PBSs) at both Alice’s and Bob’s receivers. This setup requires compensation of four fiber arms, as opposed to just one in the aforementioned reference. However, owing to the rotational invariance of the distributed entangled state, only three of these arms require active compensation.

\section{METRICS TO QUANTIFY RECEIVERS' POLARIZATION BASIS ALIGNMENT}
QBER is a commonly used metric in polarization basis alignment algorithms. In the context of the BBM92 protocol, it represents the ratio of unexpected coincidence counts to the total number of coincidence counts observed in measurement outcomes.

We denote the polarization coincidence counts as $C_{jk}$, where the indices $j,k \in \{\text{H, V, D, A}\}$ correspond to the polarization states measured at Alice’s and Bob’s receivers, respectively. This results in a total of sixteen possible polarization combinations. Assuming a stationary source that produces perfect copies of a quantum state $\ket{\Psi}$, the polarization coincidence counts are, in the simplest scenario, proportional to the projective measurement probabilities:
\begin{equation}
    C_{jk} \propto |\braket{jk|\Psi}|^2,
\end{equation}
where $\ket{jk} = \ket{j}_A \otimes \ket{k}_B$, with indices $j$ and $k$ labeling the polarization states measured at Alice's and Bob's receivers, respectively. The QBER associated with each Bell state can then be expressed in terms of the measured coincidence counts as follows:
\begin{eqnarray}
\text{QBER}_{\ket{\phi^+}} = \frac{C_{HV}+C_{VH}+C_{DA}+C_{AD}}{\beta},\nonumber \\ 
\text{QBER}_{\ket{\phi^-}} = \frac{C_{HV}+C_{VH}+C_{DD}+C_{AA}}{\beta},\nonumber \\ 
\text{QBER}_{\ket{\psi^+}} = \frac{C_{HH}+C_{VV}+C_{DA}+C_{AD}}{\beta},\nonumber \\ 
\text{QBER}_{\ket{\psi^-}} = \frac{C_{HH}+C_{VV}+C_{DD}+C_{AA}}{\beta},\nonumber  
\end{eqnarray}
where $\beta=C_{HV}+C_{VH}+C_{DA}+C_{AD} + \sum_{j\in\{\text{H,V,D,A}\} } C_{jj}$ and we label basis $Z=\{\ket{\text{H}},\ket{\text{V}}\}$ and $X=\{\ket{\text{D}}, \ket{\text{A}}\}$. Only coincidence counts measured at Alice's and Bob's detectors in the same polarization basis are used for secret key extraction and are therefore employed in the computation of QBER. The remaining eight accessible coincidence outcomes, which correspond to measurements performed in different bases, are discarded during the sifting procedure and do not contribute to the final key generation.

In the context of the polarization-based BB84 protocol with single-photon transmission, the $Z$ and $X$ bases are optimized to be mutually unbiased. This means that for any pair of basis vectors $j \in Z$ and $k \in X$, the following condition holds:
\begin{equation}
    |\braket{j|k}|^2 = 1/2      \label{critMU}
\end{equation} 
This criterion cannot be directly applied to entangled photon pairs, as the states involved in the scalar product of Eq. \eqref{critMU} must belong to the same Hilbert space. However, we can formally formulate an analogous condition for a Bell state $\ket{\Psi} \in \{\ket{\phi^+}, \ket{\phi^-}, \ket{\psi^+}, \ket{\psi^-}\}$ using the following two conditions:
\begin{align}
    &|\braket{jk|\Psi}|^2 = \frac{1}{2}|_A\braket{j|k}_A|^2 \quad\text{or} \quad|\braket{jk|\Psi}|^2 = \frac{1}{2}|_A\braket{j|\hat{X}|k}_A|^2, \label{crit1}\\
    &|\bra{jk}\hat{H}\otimes\hat{I}\ket{\Psi}|^2= |\bra{jk}\hat{I}\otimes\hat{H}\ket{\Psi}|^2=\frac{1}{4},\label{crit2}
\end{align}
These conditions are valid for all $j,k \in \{\text{H}, \text{V}\}$, and where the Pauli-$X$ and Hadamard operators are respectively defined as:
\begin{equation}
\hat{X} = 
\begin{bmatrix}
0 & 1 \\
1 & 0
\end{bmatrix}, \quad
\hat{H} = 
\frac{1}{\sqrt{2}}
\begin{bmatrix}
1 & 1 \\
1 & -1
\end{bmatrix}.
\end{equation}
If the state arriving at the receiver, $\ket{\xi}$, is maximally polarization-entangled, it can be transformed into any of the Bell states via an appropriate local unitary transformation $\hat{U}_{\text{PC}}$ implemented by polarization controllers, such that $\ket{\Psi} = \hat{U}_{\text{PC}}\ket{\xi}$. Simultaneous optimization with respect to the criteria in Eqs. \eqref{crit1} and \eqref{crit2} imposes quantum correlations at the receivers that are sufficient for QKD applications.

In a free-space receiver, the criteria in Eqs. \eqref{crit1} and \eqref{crit2} can be easily satisfied by placing waveplates in the respective optical paths and adjusting them to appropriate angles. In contrast, for a fiber-based receiver, an auxiliary laser is typically required for polarization basis alignment. Alternatively, coincidence measurements of a known entangled state can be used for this purpose, as the birefringence of single-mode fibers is unpredictable but generally stable over time in controlled environment. 

It should be noted that polarization-maintaining fibers preserve only a specific pair of orthogonal polarization states—constituting a single polarization basis—while other input bases are unitarily mixed upon transmission through the fiber.

    \begin{figure}
        \centering
        \includegraphics[width=0.570\linewidth]{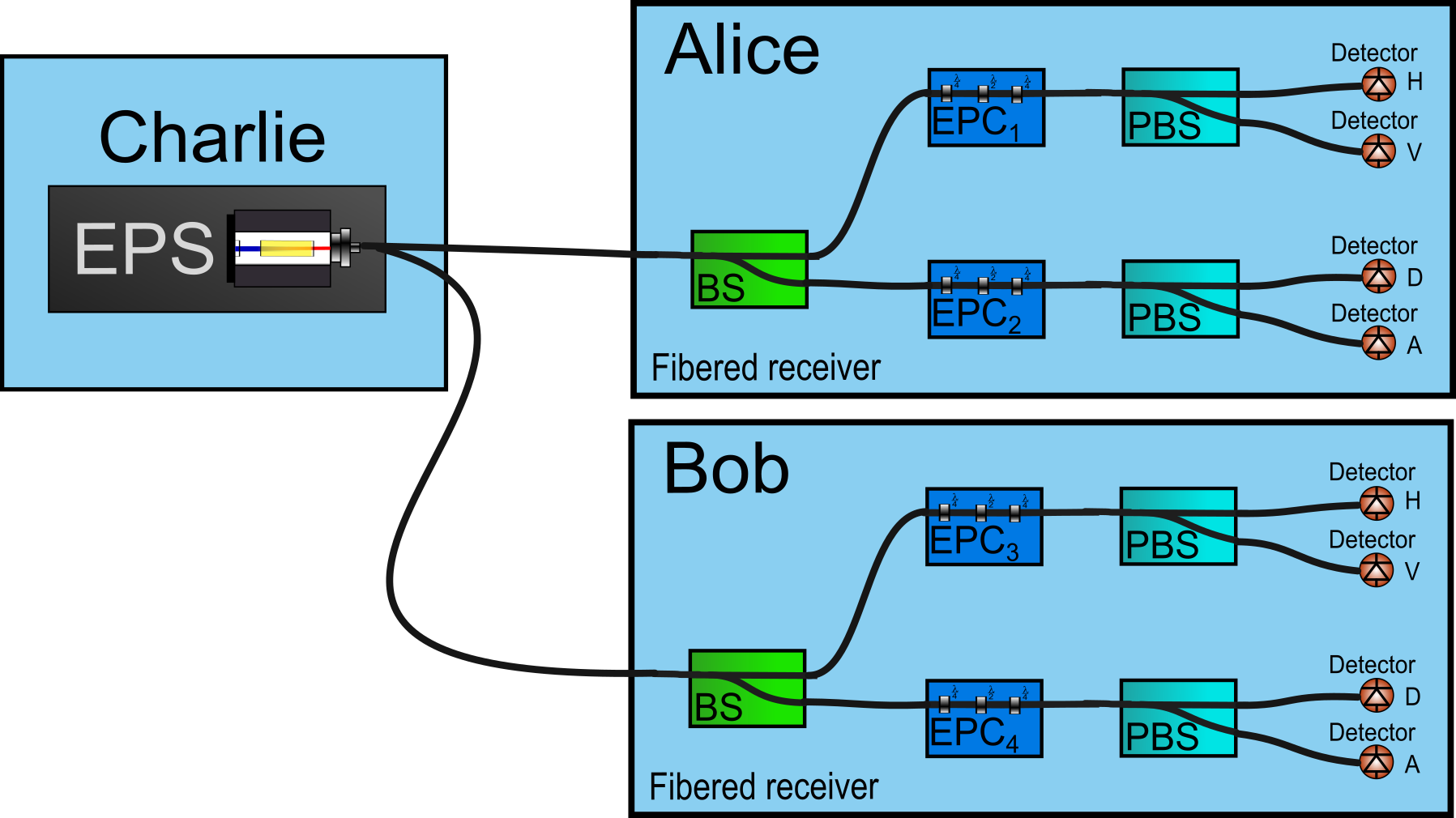}
\vspace{0.2cm}
        \caption{BBM92 laboratory testbed utilizing all-fiber optical components: an entangled photon source (EPS), beam splitters (BSs), electrically driven polarization controllers (EPCs), polarization beam splitters (PBSs), and single-photon detectors.}
        \label{fig:enter-label}
    \end{figure}

We define the coincidence entropies $H_j^{\text{same}}$ and $H_j^{\text{diff}}$ as metrics that reflect the alignment of the polarization basis in a polarization-sensitive receiver. These quantities are derived from the distribution of coincidence counts observed in the detection setup:
          \begin{equation}
              H^{\text{same}}_j = 1 - H(p_{jj},p_{jk}), \hspace{1cm} H^{\text{diff}}_j = H(p_{jl},p_{jm}),\hspace{1cm} \mathcal{H} = \sum_j H^{\text{same}}_j + H^{\text{diff}}_j, \label{eq:bilaterentr}
          \end{equation}
          where $p_{jj}=C_{jj}/N_j^{\text{same}}$, $p_{jk}=C_{jk}/N_j^{\text{same}}$, $p_{jl}=C_{jl}/N_j^{\text{diff}}$, $p_{jm}=C_{jm}/N_j^{\text{diff}}$ are probabilities, with $j\in \{H,V,D,A\}$, $k$ being a linear polarization rotated by 90$^{\circ}$ from $j$ and $l$, $m$ are different polarizations from other basis. The terms $N_j^{\text{same}} = C_{jj} + C_{jk}$, $N_j^{\text{diff}} = C_{jl} + C_{jm}$ and are chosen such, so the condition $p_1+p_2=1$ in the binary entropy function $H(p_1,p_2)=-p_1\log_2(p_1)-p_2\log_2(p_2)$ is satisfied.

    The coincidence entropy $H_j^{\text{same}}$ characterizes the distribution of coincidence counts observed at Bob's receiver, conditioned on detecting a photon with polarization $j$ at Alice's receiver, when both are measured in the same polarization basis. The entropy is computed using a binary Shannon entropy function $H(p_{jj}, p_{jk})$, which depends on the normalized coincidence counts $C_{jj}$ and $C_{jk}$. In the ideal case of perfectly preserved quantum correlations, one expects $C_{jj} \gg C_{jk}$, resulting in $H_j^{\text{same}} \approx 1$. These quantities reflect the condition described in Eq. \eqref{crit1}, which tests for the presence of strong quantum correlations in a shared measurement basis.

In contrast, the terms $H_j^{\text{diff}}$ quantify the mutual unbiasedness criterion of Eq. \eqref{crit2}, corresponding to measurements performed in different bases at Alice's and Bob's receivers. In the case of ideal basis alignment and unbiased measurement outcomes, one expects $C_{jl} \approx C_{jm}$, leading to a maximal entropy value of $H_j^{\text{diff}} \approx 1$.

As shown in Eq. \eqref{eq:bilaterentr}, the eight entropy terms $H_j^{\text{same}}$ and $H_j^{\text{diff}}$ are summed into a scalar metric $\mathcal{H}_A \leq 8$, which quantifies the polarization basis alignment of Bob's receiver. This metric is sufficient for performing basis alignment in a prepare-and-measure polarization BB84 protocol.

By evaluating Eq. \eqref{eq:bilaterentr} from Bob’s perspective—i.e., by exchanging formally the indices in the coincidence counts terms thus in probability terms—one obtains an analogous metric $\mathcal{H}_B \leq 8$, characterizing the alignment of Alice's receiver.

For simultaneous alignment of both receivers, the two metrics can be combined into a single scalar variable:
\begin{equation}
    \mathcal{H}_{\text{total}} = \mathcal{H}_A + \mathcal{H}_B,
\end{equation}
which serves as the cost function for the alignment optimization algorithm.

\section{STOCHASTIC GRADIENT DESCENT FOR POLARIZATION BASIS ALIGNMENT IN BBM92 PROTOCOL}
Our algorithm was designed for the testbed shown in Fig. \ref{fig:enter-label}, implemented for the BBM92 protocol. The setup features an entangled photon source (EPS) that spectrally separates the signal and idler photons. After propagating through a few meters of single-mode fiber, the BSs passively select the polarization basis. Electrically driven polarization controllers (EPCs) are then varied to align the polarization bases, followed by PBSs and detectors performing projective measurements.

    \begin{figure}
        \centering
        \includegraphics[width=0.5305\linewidth]{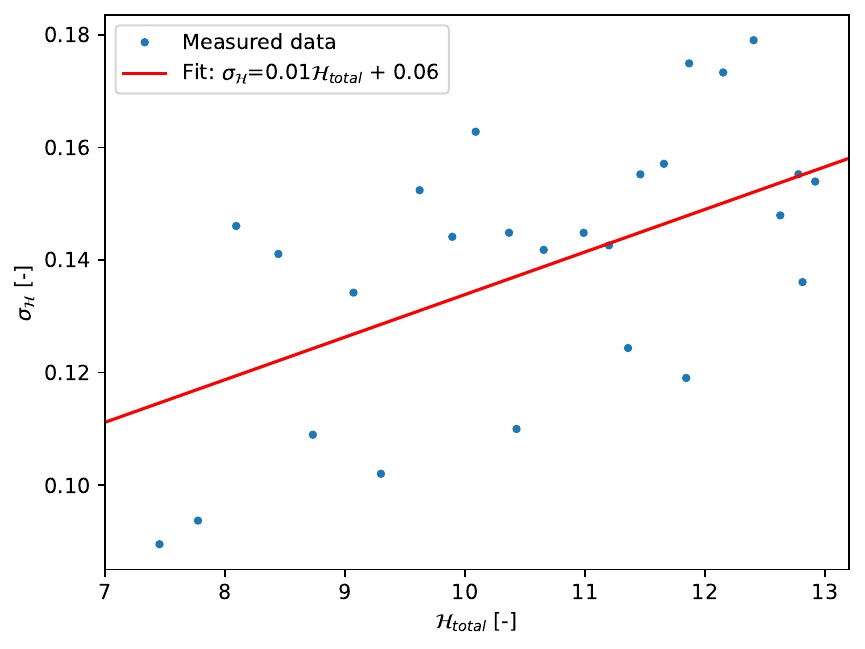}

        \caption{Dependance of standard deviation on the total coincidence entropy.}
        \label{fig:std}
    \end{figure}
Stochastic gradient descent is implemented through the following steps:
 \begin{itemize}
     \item[(1)] For each detector the single count rates are obtained at Alice $r_{A,j}$ and Bob $r_{B,j}$, where $j\in \{H,V,D,A\}$. These rates should reflect the detection efficiencies, as the preceding optical paths are identical up to a connector attenuation.
     \item[(2)] We define weights $w_{A,j}$ and $w_{B,j}$, which accounts for detection imperfections for the two detectors in same basis. The weights are given by $w_{A,j} = (r_{A,j} + r_{A,k})/r_{A,j}$, with an analogous expression for Bob. 
     \item[(3)] Coincidence counts $C_{jj}$, $C_{jk}$, $C_{jl}$, and $C_{jm}$ obtained from the detection setup are first weighted over detectors as $C_{jj}\rightarrow w_{A,j}w_{B,j}C_{jj},\dots$ and then normalized with $N_j^{\text{same}}$ and $N_j^{\text{diff}}$, resulting in coincidence probabilities $p_{jj}$, $p_{jk}$, $p_{jl}$, and $p_{jm}$.
     \item[(4)] Using the calculated probabilities, the one-sided coincidence entropies for Alice ($\mathcal{H}_A$) and Bob ($\mathcal{H}_B$) are computed. Their sum yields the two-sided coincidence entropy, defined as $\mathcal{H}_{\text{total}} = \mathcal{H}_A + \mathcal{H}_B$.
     \item[(5)] The scalar variable $\mathcal{H}_{\text{total}}$ is used as the cost function in a gradient descent optimization method, where all three paddles of each of the four EPCs are shifted from their initial positions by a step. The gradient is estimated from the change in $\mathcal{H}_{\text{total}}$ and the applied step size, and the paddles are subsequently adjusted in the direction indicated by the gradient.
    \item[(6)] Measurements are performed in 5-second intervals over $m$ trials. Increasing the number of trials reduces the standard error of the mean of the monitored variable $\mathcal{H}_{\text{total}}$.
      \end{itemize}

On Fig. \ref{fig:std} can be seen standard deviation $\sigma_{\mathcal{H}}=\sigma_{\mathcal{H}}(\mathcal{H}_{\text{total}})$ of $\mathcal{H}_{\text{total}}$ in measured 5-second intervals in sample of ten trials dependant on the value of the coincidence entropy $\mathcal{H}_{\text{total}}$. The dependence is fited with linear function with result:
\begin{equation}
    \sigma_{\mathcal{H}}(\mathcal{H_{\text{total}}}) = 0.01\mathcal{H_{\text{total}}} + 0.06.
\end{equation}
The standard error of the mean ($SEM$) can be used as a precision parameter for estimating $\mathcal{H}_{\text{total}}$, and the required number of trials can then be calculated accordingly:
\begin{equation}
m(\mathcal{H_{\text{total}}})=\left\lfloor\frac{\sigma_{\mathcal{H}}^2(\mathcal{H_{\text{total}}})}{SEM}\right\rfloor,
\end{equation}
where $\lfloor \cdot \rfloor$ denotes the floor function. This trial adaptation has not yet been implemented based on the current results. 
\begin{figure}
        \begin{subfigure}[t]{0.5\textwidth}
        \centering
        \includegraphics[width=\textwidth]{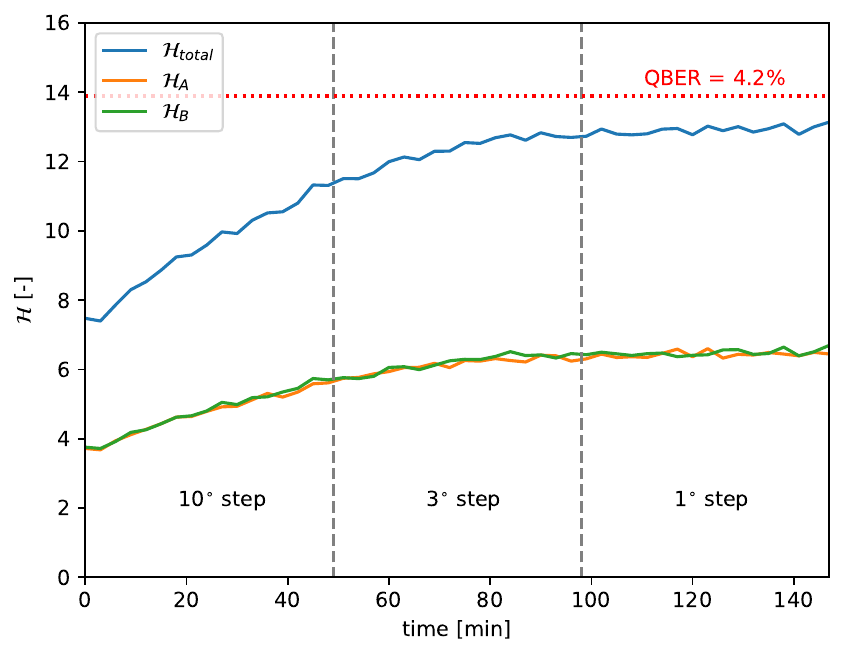}
        \caption{Evolution of coincidence entropies without averaging.}
        \label{fig:no_lines}
    \end{subfigure}
    \begin{subfigure}[t]{0.5\textwidth}
        \centering
        \includegraphics[width=\textwidth]{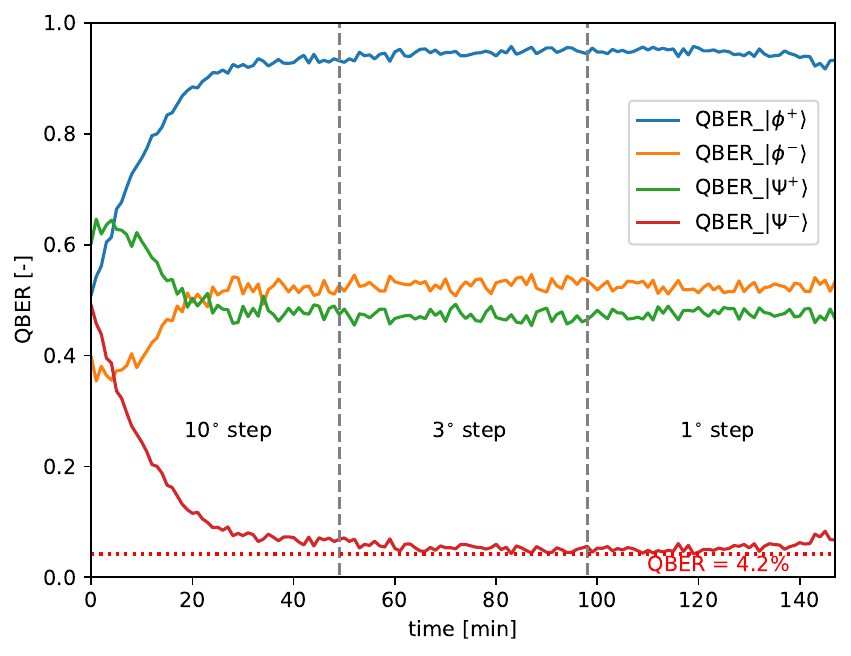}

        \caption{Evolution of QBERs without averaging.}
                        \label{fig:QBER_no}

    \end{subfigure}
      \end{figure}

The testbed shown in Fig. \ref{fig:enter-label} includes classical optical components such as BSs and PBSs. Motorized paddle polarization controllers with 18~mm diameters were used, along with single-photon avalanche diodes connected to a TimeTagger system for coincidence data acquisition. The measured coincidence rate between all polarization basis combinations of Alice and Bob was around $40~\text{s}^{-1}$, with single-photon count rates of around $6 \times 10^4~\text{s}^{-1}$.

The performance of the algorithm with $m = 1$ trial is shown in Fig. \ref{fig:no_lines} for the convergence of $\mathcal{H}_{\text{total}}$, and in Fig. \ref{fig:QBER_no} for the evolution of QBERs. No prior information about the coincidence structure of the target Bell state was used during the optimization process. The step sizes for paddle shifts in the gradient computation were varied over $10^\circ$, $5^\circ$, and $1^\circ$.

\begin{figure}
        \begin{subfigure}[t]{0.5\textwidth}
        \centering
        \includegraphics[width=\textwidth]{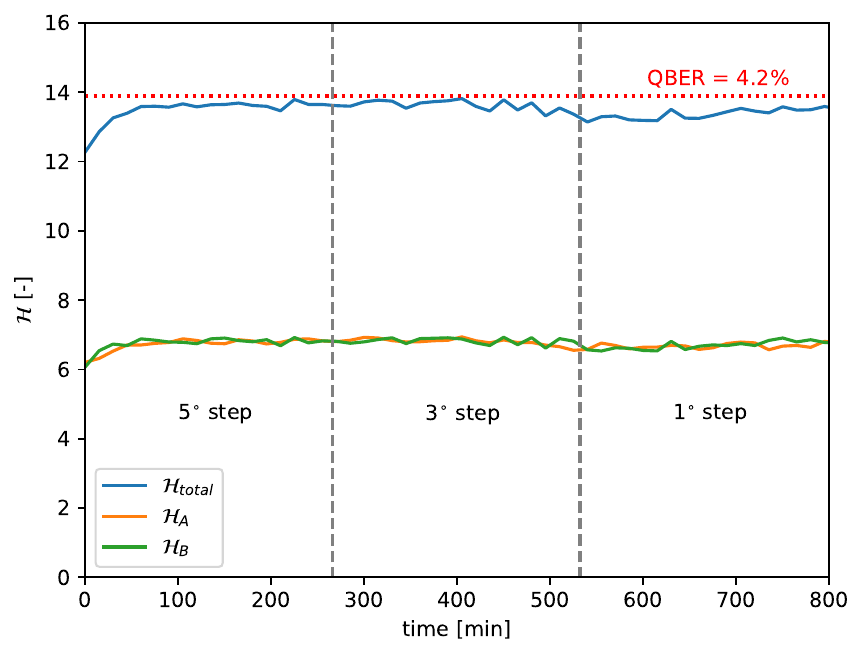}
        \caption{Evolution of coincidence entropies with averaging over five trials.}
        \label{fig:lines}

    \end{subfigure}
    \begin{subfigure}[t]{0.5\textwidth}
        \centering
        \includegraphics[width=\textwidth]{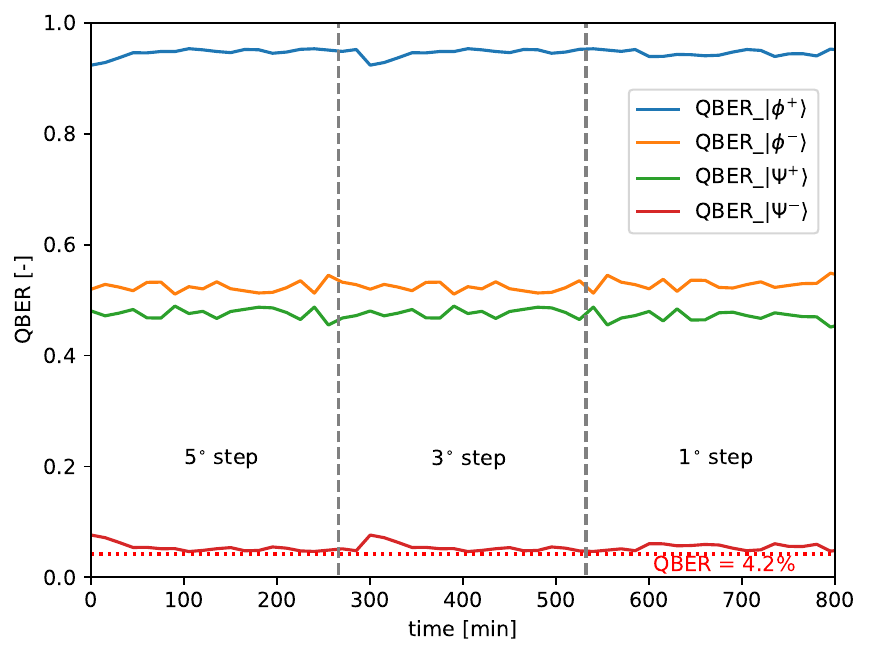}

        \caption{Evolution of QBERs with averaging over five trials.}
                \label{fig:QBER}

    \end{subfigure}
      \end{figure}
Our source produced entangled photons with an average visibility of approximately $98\%$ from both the $Z$ and $X$ bases. QBER is related to visibility $V$ via the expression $\text{QBER} = (1 - V)/2$, which implies a minimal achievable QBER of about $1\%$ in an ideal system. In our experiment, the algorithm reached a minimal value of $\text{QBER}_{\ket{\phi^+}} = 4.2\%$. 

This discrepancy may be attributed to the use of fibered EPCs and single-mode fiber transmission, which apply wavelength-dependent unitary transformations to the broadband spectrum of our EPS (with a bandwidth of approximately 60~nm centered at 1310~nm). Since the transformation matrix of the optical system is wavelength-dependent, the state of polarization at the receiver exhibits slight spectral variation.

Furthermore, we observe that $\text{QBER}_{\ket{\phi^+}}$ is closer to its theoretical minimum of 4.2\% than the corresponding coincidence entropy. This may be due to the influence of the mutual unbiasedness criterion on entropy calculations.

With averaging over five trials, the algorithm converged to higher values of total coincidence entropy, as shown in Fig. \ref{fig:lines}, while the corresponding evolution of QBERs is presented in Fig. \ref{fig:QBER}.

\section{Conclusions and Outlook}
In our approach, we define coincidence entropies to characterize the mutual unbiasedness between the polarization bases of two polarization-sensitive receivers. This scalar metric is then exploited for automated polarization basis alignment using a gradient descent algorithm.

We achieved convergence to $\text{QBER}_{\ket{\phi^+}} \sim 4.2\%$ and $\mathcal{H}_{\text{total}} \sim 13.9$ within approximately 200 minutes of algorithm operation. The primary contribution to the alignment time arises from the data acquisition required for coincidence measurements. This duration could potentially be reduced by employing a stochastic search during the initial phase of alignment or by prioritizing the optimization of the most impactful paddles.

Coincidence entropies offer an additional metric for assessing the quality of quantum communication, as they capture inter-basis correlations between the two receivers more comprehensively than QBER alone. Furthermore, coincidence counts that are typically discarded during the sifting procedure could be utilized to estimate cross-basis coincidence entropies $\mathcal{H}_{jk}^{\text{diff}}$. This may serve as a tool for transmission security monitoring or further basis adjustment.

In future work, specific QBER terms could be combined with the total coincidence entropy to form a composite cost function. When used in alignment algorithms, such a cost function could, in principle, be optimized to align the receivers with the measurement correlations of an arbitrarily chosen Bell state.

\acknowledgments 
This work was supported by The Ministry of Education, Youth, and Sports of the Czech Republic by project EH22\_008/0004649 Quantum Engineering and Nanotechnology.

\bibliography{report} 
\bibliographystyle{spiebib} 

\end{document}